# DENOISING ECG BY ADAPTIVE FILTER WITH EMPIRICAL MODE DECOMPOSITION

Bingze Dai, Wen Bai

*Abstract—* Electrocardiogram (ECG) signal is an important physiological signal which contains cardiac information and is the basis to diagnosis cardiac related diseases. In this paper, several innovative and efficient methods based on adaptive filter and empirical mode decomposition (EMD) to denoise ECG signal contaminated by various kinds of noise, including baseline wander (BW), power line interference (PLI), electrode motion artifact (EM) and muscle artifact (MA), are proposed. We first present a novel method based on EMD and adaptive filter for the removal of BW and PLI in ECG signal. We then extend the method to the complex scenario where four most common noises, PLI, BW, EM and MA are present. The proposed Parallel EMD adaptive filter structure yields the best SNR improvement on the MIT-BIH arrhythmia database, corrupted by the four types of noises.

## I. Introduction

Cardiovascular disease is one of the main threats to human life, taking an estimated 17.9 million lives each year which attributes about one third of the deaths all over the world [1]. Electrocardiogram (ECG) signal is an important basis for the diagnosis of cardiovascular disease and is used to extract information related to physiology of the heart. In practice, the ECG signals are often corrupted by several noise sources [2] such as electrode motion artifact (EM), muscle artifact (MA), power line interference (PLI), and baseline wander (BW). These noises could affect the doctor's ability to provide accurate diagnosis of cardiac function. Therefore, denoising ECG signal becomes an important task in medical and engineering fields [3]. Some traditional operations such as FIR and IIR filters [4, 5], filter bank [6], modeling [7], singular value decomposition [8] and independent component analysis [9] are used to denoise PLI, BW and white Gaussian noise. The works in [10-12] used adaptive filters to eliminate one type of noise at a time. Thakor et al [13] and Wanget et al [14] used adaptive filter-based methods and obtained promising results. There are also methods based on wavelet decomposition with various thresholding methods [15-21], extended Kalman filter [22, 23] and neural networks [24, 25] to remove the specific noise sources on ECG.

Empirical mode decomposition (EMD) [26] has been widely applied to analyze various signals, such as radar signal, vibration signal and biomedical signals including ECG. The denoising techniques based on EMD are investigated in [27-30]. EMD is also a very useful tool that can be applied with other methods. Xing et al [31] applied PCA with EMD, Singh et al [32] used method of Non-local Means and Modified EMD to denoise BW and PLI on ECG.

Although many methods have been proposed for ECG denoising, the type of noises considered are limited. In this paper, we investigate denoising algorithm for ECG signal corrupted by four types of noise: BW, PLI, EM and MA. This paper has two contributions on the denoising method. First, a novel algorithm to denoise ECG signal corrupted by PLI and BW using adaptive filter and EMD is presented. We then extend the algorithm to consider all four types of noises and propose four denoising algorithms. The algorithms are Staged Direct Adaptive Filter (SDAF), Parallel Direct Adaptive Filter (PDAF), Staged EMD Adaptive Filter (SEAF) and Parallel EMD Adaptive Filter (PEAF). The proposed algorithms are experimented on ECG signals from MIT-BIH Arrhythmia Database [33] with added noises from MIT-BIH Noise Stress Test Database [34], both databases are provided by Goldberger. A. et al [35]. The proposed methods are compared with other state-of-the-art algorithms.

The experiment result shows that the proposed EMD with adaptive filter-based method performs better to remove PLI and BW on ECG, and the proposed PEAF algorithm is the best algorithm to denoise ECG signal contaminated by BW, PLI, EM and MA. The paper is organized as follows: Section 2 provides a brief review of EMD and adaptive filter algorithms. Section 3 describes the proposed algorithm architecture designed for complex noising scenarios. The performance for each method is studied, compared and discussed with experimental results in Section 4. Section 5 concludes the paper and discusses ideas for future work.

## II. Background

EMD algorithm is proposed by Huang et al. [26] in 1998. The basic idea is to decompose the fluctuation of different scales present in the signal step by step to produce several sequences with different characteristic scales after smoothing the signal. Each sequence generated after decomposing is called an Intrinsic Mode Function (IMF). IMF is used to indicate a simple oscillator mode embedded in the data. EMD is an adaptive method that can extract IMFs of a signal according to the characteristics of each signal without any prior knowledge and the IMFs are the new bases that can represent the signal. EMD is especially well suited for analyzing nonlinear and nonstationary signals. EMD has made significant contributions in many areas including mechanical fault diagnosis, biomedical signal analysis and so on.

Adaptive filter is a digital filter that can automatically adjust the transfer function according to the input signal [36]. There are many types of adaptive filters including Least Mean Squares (LMS) filter, Recursive Least Squares (RLS) filter.

Bingze Dai is with the Department of Electrical and Computer Engineering, University of Illinois at Urbana-Champaign, Champaign, IL, 61820 USA (e-mail: daibingze@outlook.com). Wen Bai is with Institute of Engineering Mechanics, China Earthquake, Harbin, China, 150080 (e-mail: 781090853@qq.com). Wen Bai is the corresponding author.

Each method has its own algorithm and criteria to obtain the optimal adaptive filter parameters. To denoise a signal using adaptive filter, other than the noisy signal, a reference signal input which is either the signal correlated to clean signal or the signal correlated to noise is needed. Since the clean ECG is unknown, we can obtain the correlated noise signals by applying electrodes on human body and the power line. Here we use noises as reference signal. The adaptive filter structure used in the following algorithms is shown as Fig. 1. The desired signal is the noise and by subtracting it from noisy signal, denoised ECG signal is obtained.

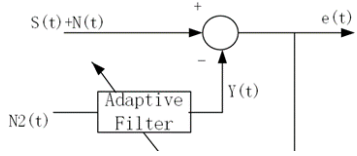

Fig. 1. Applied adaptive filter structure

### III. PROPOSED METHODS

#### A. Selection of IMF's for Denoising Each Noise

The basic idea of proposed methods is based on the property that EMD can decompose signals into stationary IMFs. Furthermore, different types of noise have their own spectrum, PLI is a single frequency noise, BW's frequency is concentrated on 0-3Hz, EM is a high frequency noise and MA occupies similar frequency range as ECG. Consequently, different noises hold different spectrum concentration which can be represent by the sum of some IMFs. Although adaptive filters are capable of coping with both stationary and non-stationary signals, the performance on stationary signals is typically better. The basic simplified structure to denoise each noise is shown in Fig. 2.

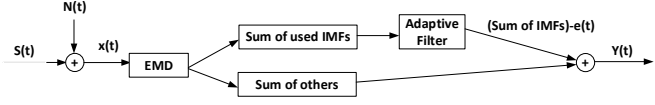

Fig. 2. Simplified denoising structure

In Fig. 2, $e(t)$ denotes the estimated noise and the output $y(t)$ denotes the denoised ECG signal, which is equal to $x(t) - e(t)$. First, the noise from MIT-BIH Noise Stress Test Database is added to the ECG signal from the MIT-BIH Arrhythmia Database. Here, clean signal plus noise $S(t) + N(t)$ is the original target signal we want to denoise. The selection is based on appling adaptive filters on IMFs' combinations. We adopt the noise related signals as the reference input signal of the adaptive filter to obtain estimated noise. The denoised combination of IMFs is obtained by subtracting the estimated noise. The denoised ECG signal is the sum of the denoised combination and other unused IMFs and residue.

By comparing the performance, which is directed by SNR improvement, $SNR_{imp}$, of using the sum of different IMFs in the adaptive filter, the best combination can be recognized. Results show that for PLI, applying the adaptive filter on the first-order IMF yields the best result whereas applying the adaptive filter on the sum of IMF5~IMF8 yields the best result for BW. For EM, IMF3~IMF8 is the best pre-denoised input summation, for MA, applying adaptive filter directly to the noised signal without EMD yields the best performance.

#### B. Methods to Denoise Different Combination of Noises

In this section, we consider the denoising problem when both BW and PLI noises corrupts the ECG signal. The proposed algorithm structure is shown in Fig. 3. Using $IMF_1'$ and $IMF_{58}'$ respectively to denote the IMF1 and IMF5~8 after adaptive filtering, the reconstructed denoised ECG signal is shown in (1):

$$Y(t) = X'(t) = IMF_1' + \sum_{k=2}^{4} IMF_k(t) + IMF_{58}' + r(t) \quad (1)$$

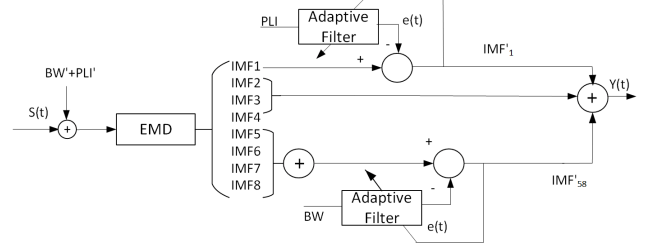

Fig. 3. Proposed algorithm structure to eliminate BW and PLI noise

When dealing with BW and PLI, there is no overlapped usage of IMFs. However, when dealing with BW, PLI, EM and MA, there are overlapping usage of IMFs. For BW, we use IMF5~8, while for EM, we use IMF3~8, the overlapping use of IMFs is a problem need to be solved. As stated above, it requires to apply adaptive filter on the whole signal to denoise MA. Therefore, the adaptive filter for denoising MA should be placed either before or after denoising the other three noises. Two methods are proposed to solve the stated problems. The first method is to process the BW and EM with a parallel structure. Corresponding adaptive filters are applied on IMF5~8 to denoise BW and IMF3~8 to denoise EM. We then sum the denoised IMF5~8 and the denoised IMF3~8 up and subtract the overlapped IMFs (IMF5~8). The second way is a staged process. We first deal with BW and get the denoised IMF5~8, denote as $IMF_{58}'$. Afterwards, IMF3 and IMF4 are added on $IMF_{58}'$ to get the adaptive filter's input to denoise EM. For denoising MA, we found that setting the adaptive filter after denoising the other three has a better performance. Based on above, we proposed PEAF and SEAF based on EMD and adaptive filter shown in Fig. 4.

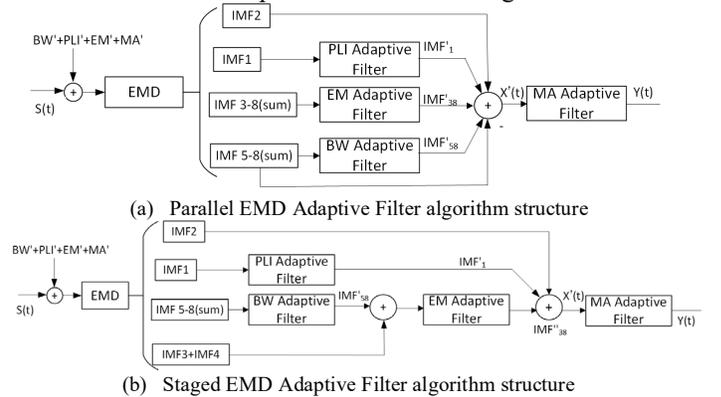

(a) Parallel EMD Adaptive Filter algorithm structure

(b) Staged EMD Adaptive Filter algorithm structure

Fig. 4. Proposed Algorithm Structure to Denoise ECG Contaminated by BW+PLI+EM+MA Based on EMD and Adaptive Filter

For PEAF algorithm, $Y(t)$, the denoised ECG, is the result of removing MA from $X'(t)$ using direct adaptive filter. $X'(t)$ is the signal after denoising PLI, BW and EM calculated as (2), where r denotes the residue after EMD.

$$X'(t) = IMF'_1 + IMF_2 + IMF'_{38} + IMF'_{58} + r - \sum_{k=5}^{8} IMF_k \quad (2)$$

For SEAF algorithm, the input signal for adaptive filter to denoise EM is the sum of $IMF3$, $IMF4$ and denoised $IMF5{\sim}8$ ($IMF'_{58}$). We use $IMF''_{38}$ to indicate the denoised $IMF3{\sim}8$ after EM adaptive filter in the SEAF. $X'(t)$ is used as the input signal of adaptive filter to denoise MA in (3).

$$X'(t) = IMF'_1 + IMF_2 + IMF''_{38} + r \quad (3)$$

To test if the EMD with adaptive filter based model has a better performance, we proposed two denoising methods based on only adaptive filter as comparison. We proposed two methods: SDAF algorithm and PDAF algorithm. SDAF refers to the method proposed by N.V. Thakor et al [13]. We add one adaptive filter layer for MA. For PDAF algorithm, we consider all types of the noises as one noise and directly use adaptive filter to denoise ECG which is the same as shown in Fig. 1. SDAF architecture is shown in Fig. 5.

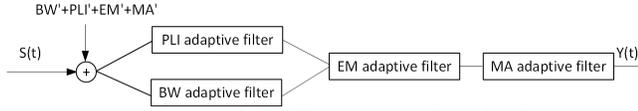

Fig. 5. Proposed SDAF structure to Denoise ECG

## IV. EXPERIMENT RESULT

In experiments, the ECG signals and BW, EM, MA added on ECG are from the MIT-BIH Database [33, 34] and PLI is a 60 Hz sinusoid. While adding the BW, EM and MA on ECG, the noises all have been shifted and passed through different filters to get noises that are not equal but correlated to the adaptive filters' reference noises. To evaluate the performance of the proposed methods, we compare them with other SOA algorithms [21,22,28]. For the first condition where ECG signals are corrupted by BW and PLI, we used SNR and MSE for evaluation. For ECG signals are contaminated by BW, PLI, EM and MA, $SNRimp$ is proposed to evaluate the performance. In (4-6), $x_c$ denotes the clean signal, $x_n$ is the noisy signal and $x_d$ is the denoised signal.

$$\text{SNR} = 10 log_{10} \left( \frac{\sum_i |x_c(i)|^2}{\sum_i |x_d(i) - x_c(i)|^2} \right) \quad (4)$$

$$RMSE = \sqrt{\frac{\sum_{i=1}^{N}(x_d(i) - x_c(i))^2}{N}} \quad (5)$$

$$\text{SNRimp} = SNR_{output} - SNR_{input}$$
$$= 10 log_{10} \left( \frac{\sum_i |x_n(i) - x_c(i)|^2}{\sum_i |x_d(i) - x_c(i)|^2} \right) \quad (6)$$

### A. BW and PLI Denoising Result

The ground truth ECG signals are randomly selected from MIT-BIH Arrhythmia Database [33]. The selected ECG signals are corrupted by BW noise from MIT-BIH Noise Stress Test Database [34] and 60Hz PLI. The corrupted ECG signals are fed into the proposed EMD and adaptive filter-based algorithm. The comparison of clean ECG, corrupted ECG and denoised ECG is shown in Fig. 6. Fig. 6(a) and (b) show the time domain signals and their frequency spectra, respectively. In Fig. 6, the first line in each group is the original clean signal, the second line is ECG signal corrupted by BW and PLI, the third line shows the denoised ECG by our method.

We can clearly see the corrupted ECG blurred lots of details in the original signal. The denoised signal is very similar to the original ECG except at the beginning because adaptive filter needs time to converge.

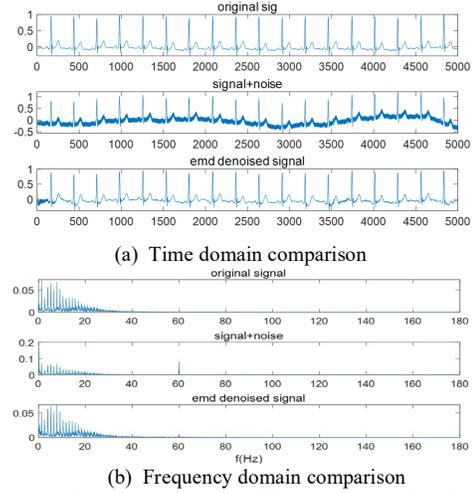

(a) Time domain comparison

(b) Frequency domain comparison

Fig. 6. Result of denoising ECG corrupted by BW+PLI

As mentioned, there are multiple algorithms for adaptive filter. Here we briefly compare LMS, NLMS and RLS using default parameters in MATLAB under same conditions to find which one is more suitable for our method. In practice, the performance of denoising is better with higher SNR and lower RMSE.

From Fig. 7 we find that LMS has a better and more stable performance at different $SNR_{input}$. As a result, we use LMS as our adaptive filter algorithm for all other methods we proposed. We compared the results of the $SNR_{output}$ under $SNR_{input} = 10$ (dB) which is commonly used in other papers. Fig. 8 compares the performance of the proposed method with other algorithms [21, 22, 28]. As evidence from the result, the proposed method has the best performance.

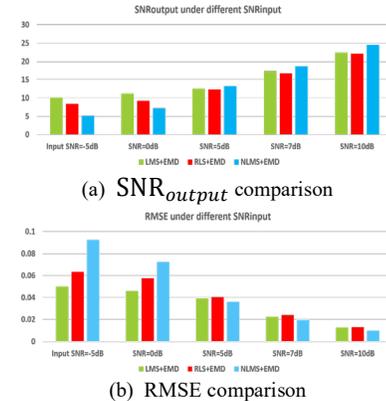

(a) $SNR_{output}$ comparison

(b) RMSE comparison

Fig. 7. Comparison of LMS NLMS and RLS under different SNRinput

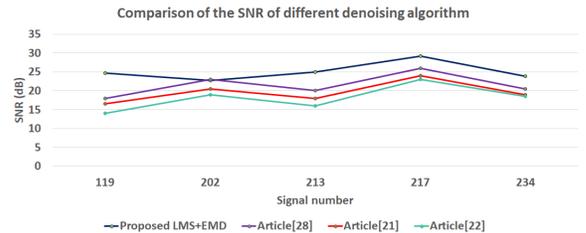

Fig. 8. Comparison with methods from Article [28,21,22]

## B. Result of Cancelling BW+PLI+EM+MA Noises

An example of ECG signal corrupted by BW, PLI, MA and EM simultaneously is shown in Fig. 9. The second line includes its time domain and spectrum, the third line shows the denoised ECG signal's time domain and spectrum. We compared four algorithms which are SDAF, PDAF, SEAF and PEAF introduced before.

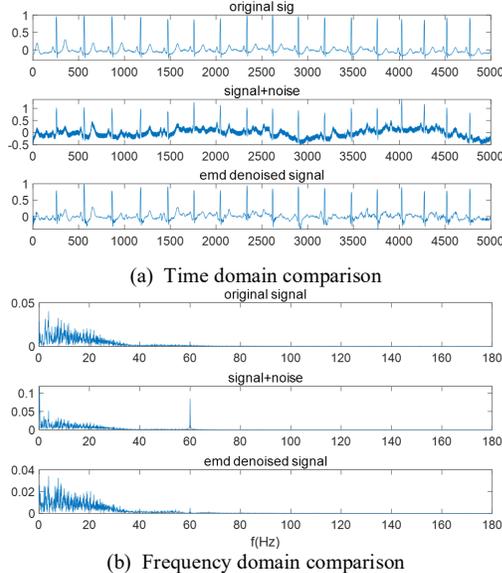

(a) Time domain comparison

(b) Frequency domain comparison

Fig. 9. Result of denoising ECG corrupted by BW+PLI+MA+EM

We evaluated the four algorithms using different ECG signals and studied the performance of each algorithm under different $SNR_{input}$. The measurement is $SNR_{imp}$, the improvement of SNR, which is the SNR difference between pre-denoise and after-denoise. By this way, we can compare the performance more directly especially under different SNR input. In Fig. 10, the four proposed methods are compared using different ECG signals and the $SNR_{imp}$ values are shown. In this case, the $SNR_{input}$ is fixed at 10dB. We found that among the four proposed methods, PEAF algorithm is observed to have the best performance in terms of $SNR_{imp}$.

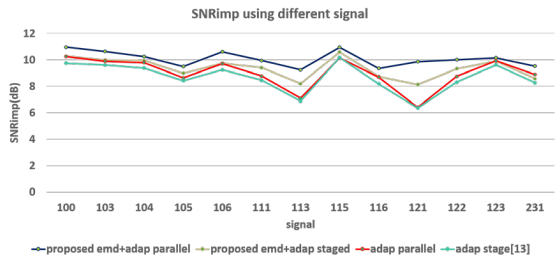

Fig. 10. Comparison of improvement in SNR with varying input SNR levels for the different denoising methods proposed in this study

Fig. 11 shows a comparison of the $SNR_{imp}$ using the proposed four methods at different $SNR_{input}$ levels. PEAF has the best performance comparing to the other three method at all levels of $SNR_{input}$.

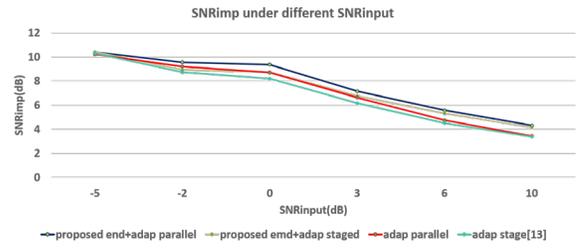

Fig. 11. Performance of four methods under different SNRinput

## V. CONCLUSION

In this paper, we proposed effective ECG denoising methods based on EMD and adaptive filter to reduce various noises. In the proposed hybrid technique, the corrupted ECG signal is first decomposed into IMFs by EMD and adaptive filters are applied on corresponding combinations of IMFs to denoise different types of noises. We first proposed a method based on EMD and adaptive filter with better performance to denoise ECG corrupted by combined noise of BW and PLI compared to previous works. Based on this idea and its better performance on cancelling BW and PLI, we proposed methods to denoise ECG corrupted by BW, PLI, MA and EM simultaneously. We compared four methods based on only adaptive filter and EMD with adaptive filter. We found that PEAF has the best performance to denoise ECG corrupted by combined noise of all four noises. A number of experiments have been carried out to demonstrate that the proposed method can serve as a good tool for denoising ECG with multiple types of noises.

In future research, we can combine different adaptive filtering algorithms with the proposed architecture in this article and study the corresponding performance more precisely. We can also consider using the proposed approach with other decomposition algorithms such as VMD [37] or EEMD [38] with adaptive filtering.

ACKNOWLEDGMENT

I would like to thank Prof. Truong Nguyen for his advice and proofreading.